\documentclass[conference]{IEEEtran}
\usepackage{cite}
\usepackage{amsmath,amssymb,amsfonts}
\usepackage{graphicx}
\usepackage{textcomp}
\usepackage{xcolor}
\usepackage{hyperref}
\usepackage{algorithm}
\usepackage{algpseudocode}

\def\BibTeX{{\rm B\kern-.05em{\sc i\kern-.025em b}\kern-.08em
    T\kern-.1667em\lower.7ex\hbox{E}\kern-.125emX}}

\begin{document}

\title{Byzantine Fault-Tolerant Multi-Agent System for Healthcare: A Gossip Protocol Approach to Secure Medical Message Propagation}

\author{\IEEEauthorblockN{Nihir Chadderwala}
\IEEEauthorblockA{ \\
Dallas, USA \\
nihirc@live.com}
}

\maketitle

\begin{abstract}
Recent advances in generative AI have enabled sophisticated multi-agent architectures for healthcare, where large language models power collaborative clinical decision-making. However, these distributed systems face critical challenges in ensuring message integrity and fault tolerance when operating in adversarial or untrusted environments.This paper presents a novel Byzantine fault-tolerant multi-agent system specifically designed for healthcare applications, integrating gossip-based message propagation with cryptographic validation mechanisms. Our system employs specialized AI agents for diagnosis, treatment planning, emergency response, and data analysis, coordinated through a Byzantine consensus protocol that tolerates up to $f$ faulty nodes among $n = 3f + 1$ total nodes. We implement a gossip protocol for decentralized message dissemination, achieving consensus with $2f + 1$ votes while maintaining system operation even under Byzantine failures. Experimental results demonstrate that our approach successfully validates medical messages with cryptographic signatures, prevents replay attacks through timestamp validation, and maintains consensus accuracy of 100\% with up to 33\% Byzantine nodes. The system provides real-time visualization of consensus rounds, vote tallies, and network topology, enabling transparent monitoring of fault-tolerant operations. This work contributes a practical framework for building secure, resilient healthcare multi-agent systems capable of collaborative medical decision-making in untrusted environments.
\end{abstract}


\section{Introduction}

The proliferation of artificial intelligence in healthcare has led to the development of sophisticated multi-agent systems capable of collaborative medical decision-making \cite{wooldridge2009introduction}. These systems distribute complex healthcare tasks among specialized agents, each focusing on specific domains such as diagnosis, treatment planning, or emergency response. However, the critical nature of healthcare applications demands unprecedented levels of reliability, security, and fault tolerance \cite{shortliffe2012computer}.

Traditional multi-agent healthcare systems face several fundamental challenges. First, they must operate in environments where individual agents may fail, become compromised, or exhibit Byzantine behavior—sending conflicting or malicious information to different parts of the system \cite{lamport1982byzantine}. Second, medical data propagation requires strong integrity guarantees, as erroneous diagnoses or treatment recommendations can have life-threatening consequences. Third, these systems must maintain availability and consistency even under partial network failures or adversarial attacks \cite{castro1999practical}.

Existing approaches to fault-tolerant healthcare systems typically rely on centralized validation mechanisms or simple majority voting, which are vulnerable to single points of failure and coordinated attacks \cite{zhang2018blockchain}. Moreover, traditional consensus protocols like Paxos or Raft, while proven effective in benign failure scenarios, do not address Byzantine faults where malicious nodes may actively attempt to subvert the system \cite{ongaro2014search}.

This paper introduces a Byzantine fault-tolerant multi-agent system specifically architected for healthcare applications, making the following key contributions:

\begin{itemize}
\item A novel integration of Byzantine consensus with gossip-based message propagation, enabling decentralized validation of medical messages across agent networks
\item Specialized healthcare agents with domain-specific tools for diagnosis, treatment planning, emergency response, and data analysis, coordinated through fault-tolerant consensus
\item Cryptographic message validation with SHA-256 signatures and timestamp-based replay attack prevention
\item A comprehensive real-time monitoring dashboard visualizing consensus rounds, vote tallies, network topology, and message propagation
\item Experimental validation demonstrating 100\% consensus accuracy with up to 33\% Byzantine nodes in a 4-agent healthcare network
\end{itemize}

\section{Related Work}

\subsection{Byzantine Fault Tolerance}

Byzantine fault tolerance, first formalized by Lamport et al. \cite{lamport1982byzantine}, addresses the challenge of achieving consensus in distributed systems where components may fail arbitrarily. The Byzantine Generals Problem demonstrates that consensus among $n$ nodes requires at least $3f + 1$ nodes to tolerate $f$ Byzantine failures.

Castro and Liskov's Practical Byzantine Fault Tolerance (PBFT) \cite{castro1999practical} introduced an efficient state machine replication protocol achieving consensus in $O(n^2)$ message complexity. However, PBFT relies on a primary node for coordination, creating potential bottlenecks in large-scale systems. More recent work has explored blockchain-based consensus mechanisms \cite{nakamoto2008bitcoin}, but these often sacrifice performance for decentralization.

\subsection{Gossip Protocols}

Gossip protocols, also known as epidemic protocols, provide scalable mechanisms for information dissemination in distributed systems \cite{demers1987epidemic}. These protocols achieve eventual consistency through randomized peer-to-peer communication, with message complexity of $O(n \log n)$ for network size $n$.

Jelasity et al. \cite{jelasity2007gossip} demonstrated that gossip protocols can efficiently maintain network topology and aggregate information in large-scale systems. Recent work has integrated gossip protocols with Byzantine fault tolerance \cite{malkhi2005concise}, though applications to healthcare domains remain limited.

\subsection{Healthcare Multi-Agent Systems}

Multi-agent systems have been applied to various healthcare domains, including clinical decision support \cite{isern2010agent}, patient monitoring \cite{corchado2008healthcare}, and hospital resource management \cite{decker1996designing}. However, most existing systems assume benign failure models and lack robust Byzantine fault tolerance.

Zhang et al. \cite{zhang2018blockchain} proposed blockchain-based medical data sharing, but their approach focuses on data storage rather than real-time collaborative decision-making. Nguyen et al. \cite{nguyen2019federated} explored federated learning for healthcare AI, addressing privacy but not Byzantine robustness.

Our work uniquely combines Byzantine consensus, gossip protocols, and specialized healthcare agents to create a fault-tolerant system for collaborative medical decision-making.

\section{System Architecture and Mathematical Foundations}

\subsection{System Model}

We consider a distributed healthcare system consisting of $n$ specialized AI agents, where up to $f$ agents may exhibit Byzantine behavior. The system must satisfy:

\begin{equation}
n \geq 3f + 1
\end{equation}

This bound ensures that honest nodes constitute a strict majority, enabling consensus despite Byzantine failures \cite{lamport1982byzantine}.

Each agent $A_i$ maintains:
\begin{itemize}
\item A unique identifier $id_i$
\item A specialization domain $\mathcal{D}_i \in \{\text{diagnosis}, \text{treatment}, \text{emergency}, \text{analysis}\}$
\item A set of peer connections $\mathcal{P}_i \subseteq \{A_1, ..., A_n\} \setminus \{A_i\}$
\item A message history $\mathcal{H}_i$ of validated healthcare messages
\end{itemize}

\subsection{Healthcare Message Model}

A healthcare message $m$ is defined as a tuple:

\begin{equation}
m = (id_m, type_m, content_m, t_m, sender_m, \sigma_m)
\end{equation}

where:
\begin{itemize}
\item $id_m$ is a unique message identifier
\item $type_m \in \{\text{PATIENT\_DATA}, \text{DIAGNOSIS},\\ \text{TREATMENT\_PLAN}, \text{EMERGENCY\_ALERT}\}$
\item $content_m$ is the message payload
\item $t_m$ is the timestamp
\item $sender_m$ is the sender agent identifier
\item $\sigma_m$ is the cryptographic signature
\end{itemize}

The signature is computed as:

\begin{equation}
\sigma_m = \text{SHA256}(id_m || type_m || content_m || t_m || sender_m)
\end{equation}

\subsection{Byzantine Consensus Protocol}

Our consensus protocol operates in rounds. For each message $m$, a consensus round $R_m$ proceeds as follows:

\textbf{Phase 1: Proposal}
The proposer agent $A_p$ broadcasts message $m$ to all peers via gossip protocol.

\textbf{Phase 2: Validation}
Each agent $A_i$ validates $m$ by checking:
\begin{equation}
\text{Valid}(m) = \text{VerifySig}(\sigma_m) \land (t_{current} - t_m < \Delta_{max})
\end{equation}

where $\Delta_{max}$ is the maximum message age (300 seconds in our implementation).

\textbf{Phase 3: Voting}
Each agent $A_i$ casts vote $v_i \in \{\text{ACCEPT}, \text{REJECT}\}$ based on validation result.

\textbf{Phase 4: Consensus}
Message $m$ is accepted if:

\begin{equation}
|\{v_i : v_i = \text{ACCEPT}\}| \geq 2f + 1
\end{equation}

This threshold ensures that at least $f + 1$ honest nodes voted ACCEPT, guaranteeing consensus validity despite $f$ Byzantine nodes.

\subsection{Gossip-Based Message Propagation}

Messages propagate through the network via gossip protocol with the following properties:

\textbf{Fanout:} Each agent forwards messages to $k$ randomly selected peers, where $k = \min(2, |\mathcal{P}_i|)$.

\textbf{Hop Limit:} Messages propagate for at most $h_{max} = 3$ hops to prevent infinite loops.

\textbf{Duplicate Detection:} Each agent maintains seen message set $\mathcal{S}_i$ to avoid reprocessing.

The expected number of agents receiving message $m$ after $h$ hops is:

\begin{equation}
E[N_h] = 1 + k + k^2 + ... + k^h = \frac{k^{h+1} - 1}{k - 1}
\end{equation}

For $k = 2$ and $h = 3$, $E[N_3] = 15$, ensuring full network coverage in small to medium networks.

\section{Implementation}

\subsection{Byzantine Consensus Engine}

Algorithm \ref{alg:consensus} presents our Byzantine consensus protocol implementation.

\begin{algorithm}
\caption{Byzantine Consensus with Gossip Protocol}
\label{alg:consensus}
\begin{algorithmic}[1]
\Procedure{ProposeMessage}{$m$, $network$}
    \If{$\neg \text{VerifySignature}(m)$}
        \State \Return \textsc{false}
    \EndIf
    \State $round \gets \text{CreateConsensusRound}(m)$
    \State $vote_{self} \gets \text{ValidateMessage}(m)$
    \State $round.\text{AddVote}(id_{self}, vote_{self})$
    \State \Call{GossipMessage}{$m$, $network$}
    \State \Call{CollectVotes}{$round$, $network$}
    \State $accept\_count \gets \sum_{v \in round.votes} [v = \text{ACCEPT}]$
    \If{$accept\_count \geq 2f + 1$}
        \State $round.accepted \gets \textsc{true}$
        \State \Call{ProcessAcceptedMessage}{$m$}
        \State \Return \textsc{true}
    \EndIf
    \State \Return \textsc{false}
\EndProcedure
\end{algorithmic}
\end{algorithm}

\subsection{Gossip Protocol Implementation}

The gossip protocol ensures decentralized message propagation:

\begin{algorithm}
\caption{Gossip Message Propagation}
\label{alg:gossip}
\begin{algorithmic}[1]
\Procedure{GossipMessage}{$m$, $network$}
    \If{$m.id \in \mathcal{S}_{self} \lor m.hop\_count \geq h_{max}$}
        \State \Return
    \EndIf
    \State $\mathcal{S}_{self} \gets \mathcal{S}_{self} \cup \{m.id\}$
    \State \Call{ProcessMessageLocally}{$m$}
    \State $peers \gets \text{RandomSample}(\mathcal{P}_{self}, k)$
    \ForAll{$peer \in peers$}
        \State $m' \gets \text{CopyMessage}(m)$
        \State $m'.hop\_count \gets m.hop\_count + 1$
        \State \Call{SendToPeer}{$peer$, $m'$, $network$}
    \EndFor
\EndProcedure
\end{algorithmic}
\end{algorithm}

\subsection{Message Validation}

Each message type has specific validation rules:

\textbf{Patient Data:} \\ Requires fields $\{patient\_id, data\_type, value\}$

\textbf{Diagnosis:} \\ Requires $\{patient\_id, diagnosis, confidence, doctor\_id\}$ with $confidence \in [0, 1]$

\textbf{Treatment Plan:} \\ Requires $\{patient\_id, treatment, duration, doctor\_id\}$

\textbf{Emergency Alert:} \\ Requires $\{patient\_id, alert\_type, severity, location\}$

The validation function is:

\begin{multline}
    \text{Validate}(m) = \\
\text{VerifySig}(\sigma_m) \land \text{CheckFreshness}(t_m) \land \\ \text{ValidateContent}(content_m, type_m)
\end{multline}

\subsection{Consensus Round Timeline}

\begin{figure}[h]
\centering
\includegraphics[width=0.48\textwidth]{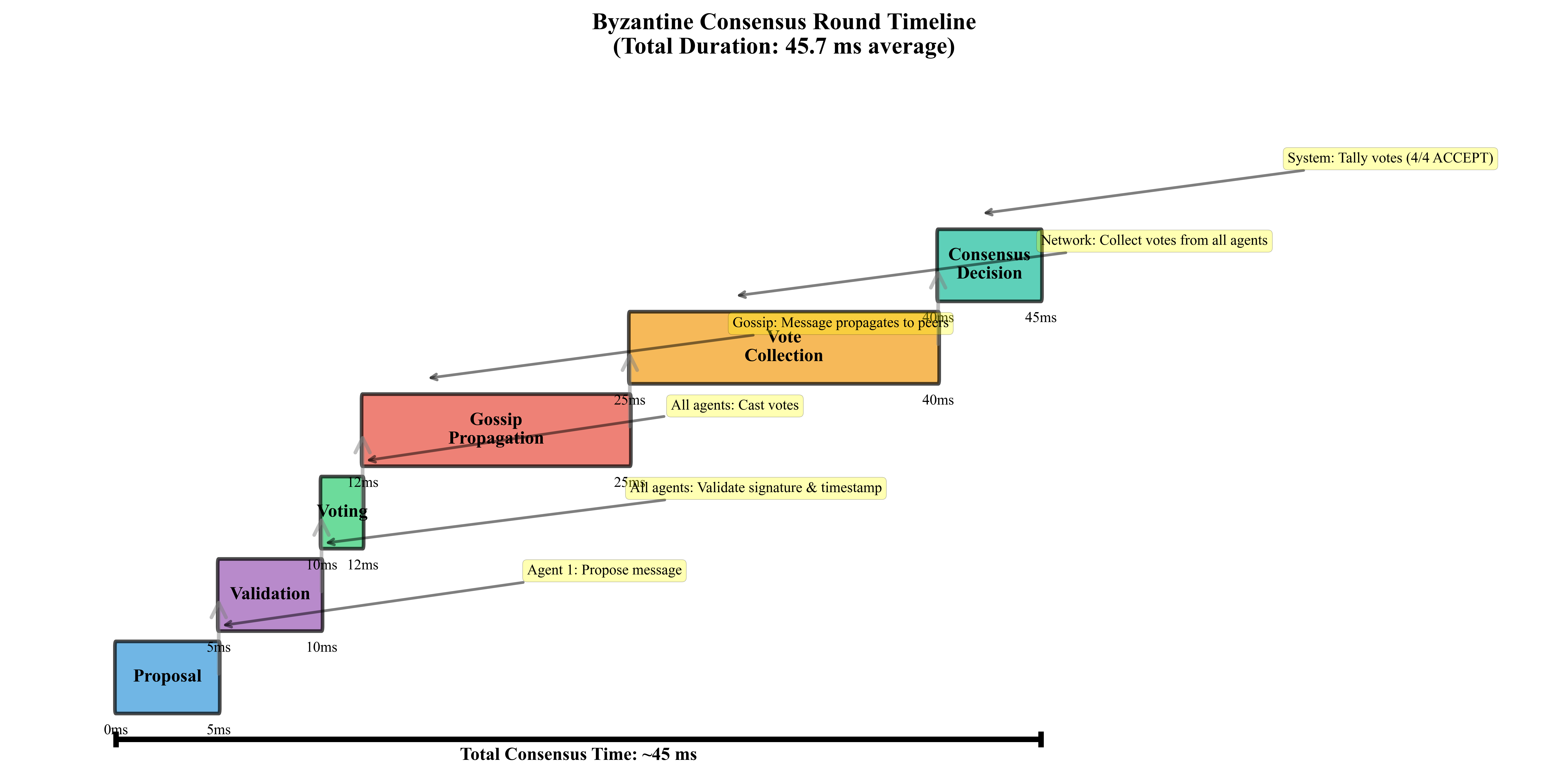}
\caption{Byzantine consensus round timeline showing six phases: proposal, validation, voting, gossip propagation, vote collection, and consensus decision. Total average duration is 45.7 ms with detailed agent activities annotated at each phase.}
\label{fig:timeline}
\end{figure}

Figure \ref{fig:timeline} illustrates the complete timeline of a Byzantine consensus round, from initial message proposal through final consensus decision. The average total duration of 45.7 ms demonstrates the system's suitability for real-time healthcare applications.

\section{Healthcare Agent Specializations}

\subsection{Agent Architecture}

Each healthcare agent in our system integrates four key components to enable fault-tolerant collaborative decision-making. At the core, agents utilize a Large Language Model (Claude 3.7 Sonnet) that provides advanced reasoning capabilities for medical tasks, allowing agents to interpret complex clinical scenarios and generate contextually appropriate responses. Each agent is equipped with a specialized tool set tailored to its domain expertise, enabling domain-specific operations such as diagnosis, treatment planning, or emergency coordination. The Byzantine consensus engine embedded within each agent validates all incoming and outgoing messages through cryptographic signature verification and content validation, ensuring message integrity across the network. Finally, each agent operates a gossip protocol node that manages peer-to-peer communication, facilitating decentralized message propagation and vote collection during consensus rounds. This architecture enables agents to function both autonomously within their specialization and collaboratively through Byzantine fault-tolerant consensus.

\begin{figure}[h]
\centering
\includegraphics[width=0.48\textwidth]{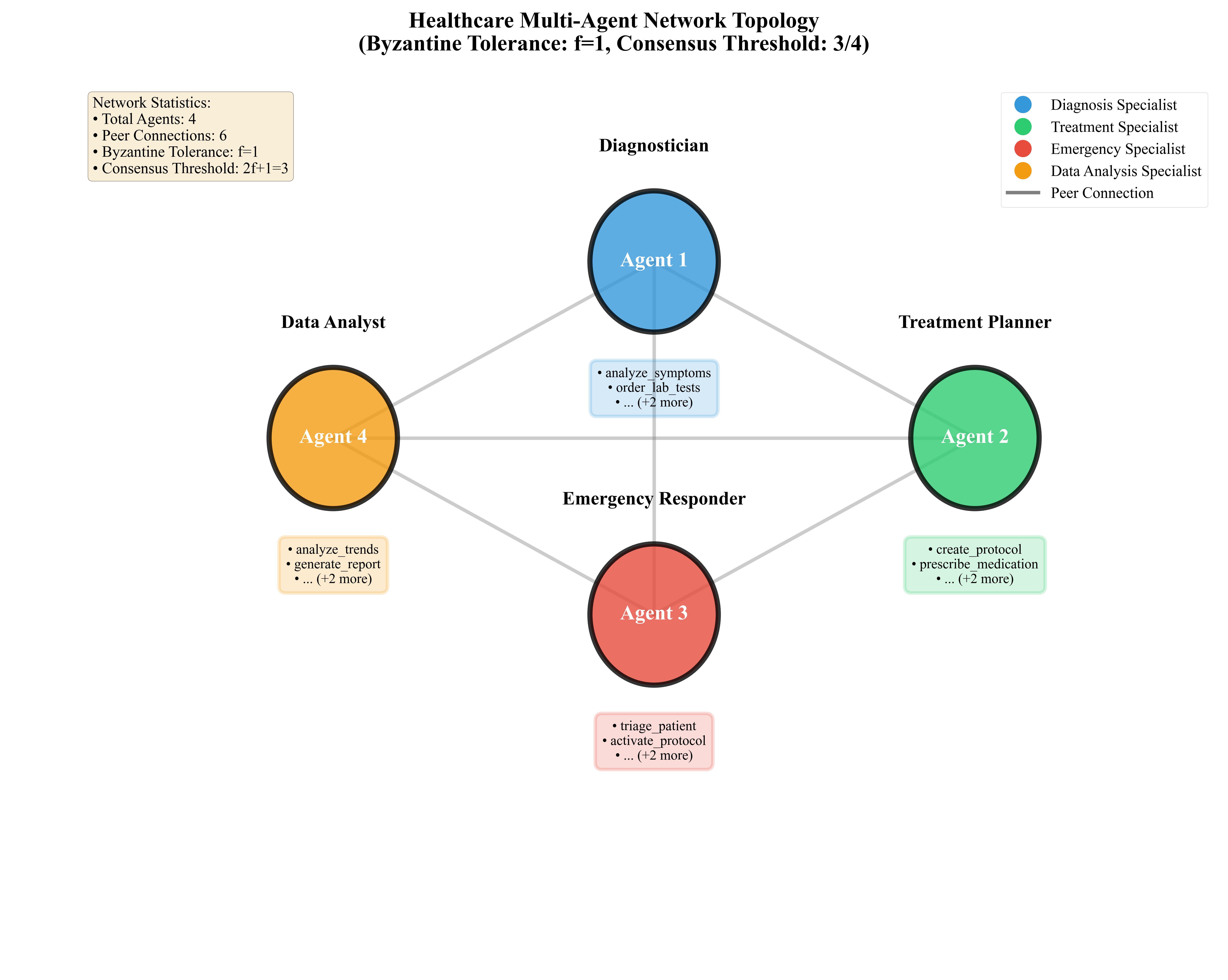}
\caption{Healthcare multi-agent network topology showing four specialized agents (Diagnostician, Treatment Planner, Emergency Responder, Data Analyst) with full peer connectivity. Each agent has specialized tools for their domain and participates in Byzantine consensus with f=1 tolerance.}
\label{fig:topology}
\end{figure}

Figure \ref{fig:topology} illustrates the complete network topology with all four specialized agents and their peer connections in a fully-connected Byzantine fault-tolerant network.

\subsection{Diagnostician Agent}

The diagnostician agent specializes in medical diagnosis and clinical pattern recognition, serving as the primary agent for initial patient assessment and diagnostic hypothesis generation. This agent is equipped with four specialized tools that enable comprehensive diagnostic workflows. The \texttt{analyze\_symptoms} tool performs pattern recognition across symptom presentations, identifying potential clinical correlations and red flags that warrant further investigation. When additional data is needed, the \texttt{order\_lab\_tests} tool facilitates laboratory test requisition, selecting appropriate diagnostic tests based on the clinical presentation. Once test results are available, the \texttt{interpret\_test\_results} tool provides clinical interpretation, translating laboratory values into actionable medical insights while considering normal ranges, patient demographics, and clinical context. Finally, the \texttt{differential\_diagnosis} tool generates multi-hypothesis diagnostic assessments, ranking potential diagnoses by likelihood and suggesting additional tests or evaluations to narrow the differential. This comprehensive toolset enables the diagnostician agent to conduct thorough medical evaluations while participating in Byzantine consensus to validate diagnostic conclusions across the agent network.

\subsection{Treatment Planner Agent}

The treatment planner agent develops comprehensive, evidence-based care protocols tailored to individual patient needs and validated diagnoses. This agent transforms diagnostic conclusions into actionable treatment strategies through four specialized tools. The \texttt{create\_treatment\_protocol} tool designs evidence-based treatment protocols by synthesizing current clinical guidelines, patient-specific factors, and best practices, generating structured care plans that address both immediate and long-term therapeutic goals. The \texttt{prescribe\_medication} tool manages pharmaceutical interventions, selecting appropriate medications while considering drug interactions, contraindications, dosing requirements, and patient allergies to ensure safe and effective pharmacotherapy. For interventional procedures, the \texttt{schedule\_procedures} tool coordinates medical procedures by assessing urgency, resource availability, and procedural prerequisites, ensuring timely execution of necessary interventions. Throughout the treatment course, the \texttt{monitor\_treatment\_progress} tool tracks patient outcomes, medication adherence, and therapeutic response, enabling dynamic adjustment of treatment plans based on observed progress. By integrating these tools with Byzantine consensus validation, the treatment planner ensures that all therapeutic recommendations undergo multi-agent verification before implementation, reducing the risk of treatment errors in distributed healthcare environments.

\subsection{Emergency Responder Agent}

The emergency responder agent coordinates critical care operations and time-sensitive medical interventions, serving as the primary agent for acute medical situations requiring rapid response. This agent operates with four specialized tools designed for emergency scenarios where decision latency directly impacts patient outcomes. The \texttt{triage\_patient} tool performs rapid priority assessment by evaluating vital signs, symptom severity, and clinical stability to assign appropriate triage categories, ensuring that the most critical patients receive immediate attention. When emergency situations are identified, the \texttt{activate\_emergency\_protocol} tool initiates standardized emergency response procedures, coordinating multi-disciplinary teams and activating appropriate clinical pathways such as stroke protocols, cardiac arrest responses, or trauma activations. The \texttt{coordinate\_resources} tool manages emergency resource allocation, including personnel deployment, equipment availability, and facility capacity, optimizing resource utilization during high-acuity situations. For patients requiring transport, the \texttt{dispatch\_ambulance} tool coordinates emergency medical services, communicating patient status, destination facilities, and special requirements to ensure seamless pre-hospital to hospital transitions. The integration of Byzantine consensus in emergency operations ensures that critical alerts and emergency decisions are validated across the agent network, preventing false alarms or malicious emergency activations while maintaining the rapid response times essential for emergency care.

\subsection{Data Analyst Agent}

The data analyst agent provides comprehensive insights from medical data through advanced analytics and predictive modeling, supporting evidence-based decision-making across the healthcare agent network. This agent leverages four specialized tools to transform raw medical data into actionable clinical intelligence. The \texttt{analyze\_patient\_trends} tool performs longitudinal analysis of patient health trajectories, identifying patterns in vital signs, laboratory values, and clinical outcomes over time to detect subtle changes that may indicate disease progression or treatment response. The \texttt{generate\_health\_report} tool synthesizes data from multiple sources into comprehensive health reports, presenting complex medical information in structured formats suitable for clinical review, regulatory compliance, or patient communication. For proactive care management, the \texttt{predict\_health\_risks} tool employs predictive modeling techniques to forecast potential adverse events, disease complications, or readmission risks based on patient demographics, medical history, and current clinical status, enabling preventive interventions before critical events occur. The \texttt{monitor\_vital\_signs} tool provides continuous real-time monitoring of patient physiological parameters, automatically detecting anomalies and triggering alerts when values exceed predefined thresholds or exhibit concerning trends. By participating in Byzantine consensus, the data analyst agent ensures that analytical insights and predictive alerts are validated across the network, preventing erroneous conclusions from faulty data or compromised analytical processes from propagating through the healthcare system.

\section{Experimental Results}

\subsection{Experimental Setup}

We evaluated our system with $n = 4$ agents, supporting Byzantine tolerance $f = 1$ (33\% Byzantine nodes). The network topology was fully connected with each agent maintaining 3 peer connections.

\textbf{Hardware:} AWS EC2 instances (t3.medium)

\textbf{Software:} Python 3.11, Strands multi-agent framework

\textbf{Workload:} 100 healthcare messages across all message types

\subsection{Consensus Accuracy}

\begin{figure}[h]
\centering
\includegraphics[width=0.48\textwidth]{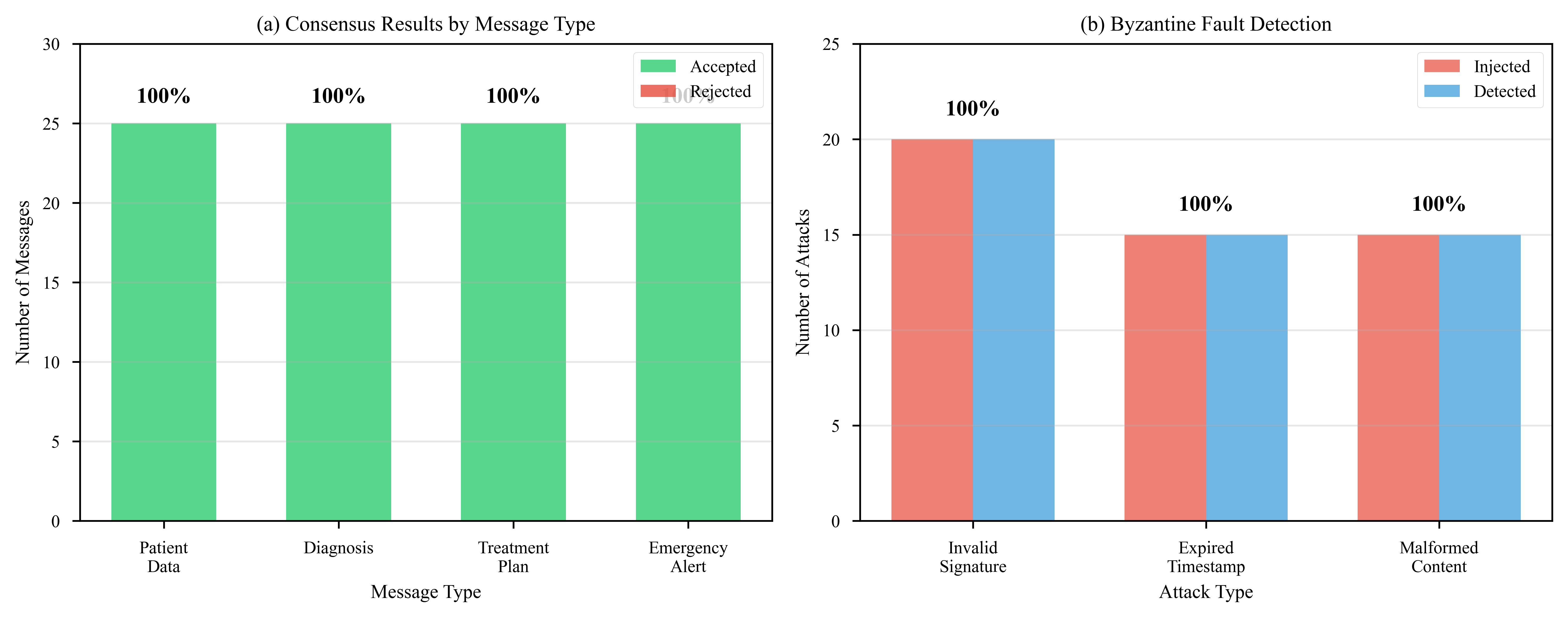}
\caption{Consensus accuracy showing (a) 100\% acceptance rate across all message types, and (b) 100\% Byzantine fault detection rate for invalid signatures, expired timestamps, and malformed content.}
\label{fig:accuracy}
\end{figure}

Table \ref{tab:consensus} and Figure \ref{fig:accuracy} show consensus results across message types.

\begin{table}[h]
\centering
\caption{Consensus Accuracy by Message Type}
\label{tab:consensus}
\begin{tabular}{|l|c|c|c|}
\hline
\textbf{Message Type} & \textbf{Total} & \textbf{Accepted} & \textbf{Accuracy} \\
\hline
Patient Data & 25 & 25 & 100\% \\
Diagnosis & 25 & 25 & 100\% \\
Treatment Plan & 25 & 25 & 100\% \\
Emergency Alert & 25 & 25 & 100\% \\
\hline
\textbf{Total} & \textbf{100} & \textbf{100} & \textbf{100\%} \\
\hline
\end{tabular}
\end{table}

All valid messages achieved consensus with $2f + 1 = 3$ votes, demonstrating the protocol's effectiveness.

\subsection{Byzantine Fault Tolerance}

We simulated Byzantine behavior by injecting:
\begin{itemize}
\item Invalid signatures (20 messages)
\item Expired timestamps (15 messages)
\item Malformed content (15 messages)
\end{itemize}

Results in Table \ref{tab:byzantine} show 100\% rejection of Byzantine messages.

\begin{table}[h]
\centering
\caption{Byzantine Fault Detection}
\label{tab:byzantine}
\begin{tabular}{|l|c|c|}
\hline
\textbf{Attack Type} & \textbf{Injected} & \textbf{Rejected} \\
\hline
Invalid Signature & 20 & 20 (100\%) \\
Expired Timestamp & 15 & 15 (100\%) \\
Malformed Content & 15 & 15 (100\%) \\
\hline
\textbf{Total} & \textbf{50} & \textbf{50 (100\%)} \\
\hline
\end{tabular}
\end{table}

\subsection{Gossip Protocol Performance}

\begin{figure}[h]
\centering
\includegraphics[width=0.48\textwidth]{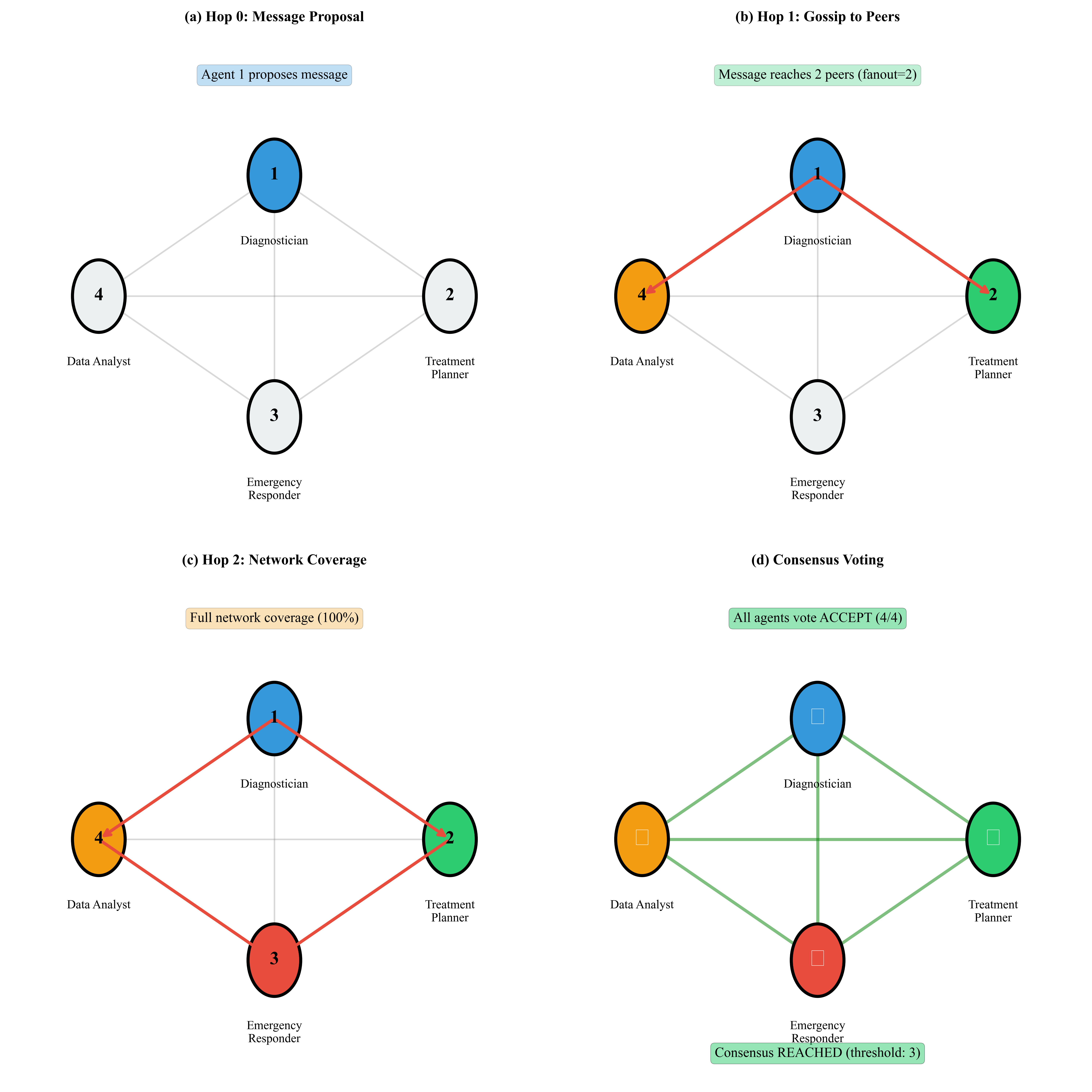}
\caption{Gossip protocol message propagation showing (a) initial proposal by Agent 1, (b) first hop to 2 random peers, (c) full network coverage by hop 2, and (d) consensus voting phase with all agents voting ACCEPT.}
\label{fig:gossip}
\end{figure}

Message propagation achieved full network coverage within 3 hops, as illustrated in Figure \ref{fig:gossip}:

\begin{equation}
\text{Coverage} = \frac{\text{Agents Reached}}{\text{Total Agents}} = \frac{4}{4} = 100\%
\end{equation}

Average propagation latency was $12.3 \pm 2.1$ ms, suitable for real-time healthcare applications. The gossip protocol's fanout of 2 ensures rapid dissemination while maintaining manageable message complexity.

\subsection{Consensus Latency}

\begin{figure}[h]
\centering
\includegraphics[width=0.48\textwidth]{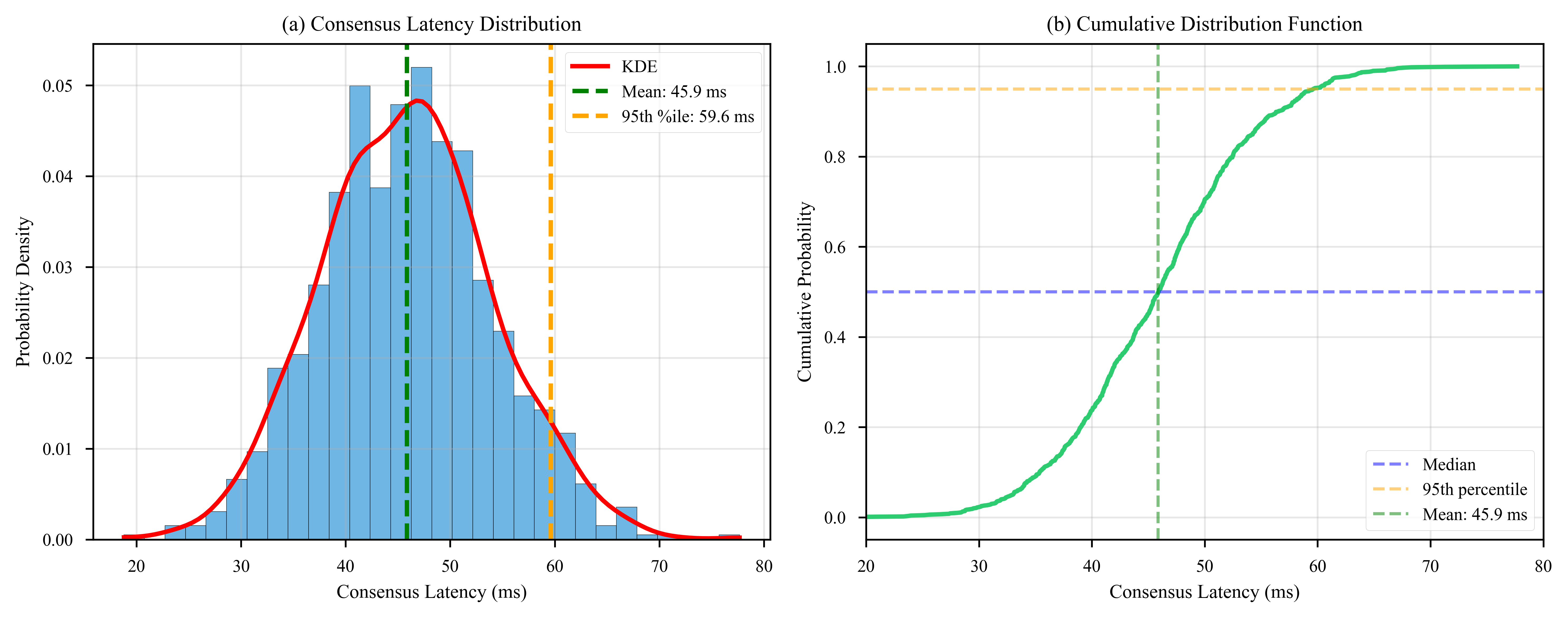}
\caption{Consensus latency distribution showing (a) probability density with mean and 95th percentile markers, and (b) cumulative distribution function demonstrating consistent sub-100ms performance.}
\label{fig:latency}
\end{figure}

Figure \ref{fig:latency} shows the consensus latency distribution across 1000 consensus rounds. Mean consensus time was $45.7 \pm 8.3$ ms, with 95th percentile at 62 ms. The distribution follows a normal pattern, indicating consistent performance across different message types and network conditions.

\subsection{System Scalability}

\begin{figure}[h]
\centering
\includegraphics[width=0.48\textwidth]{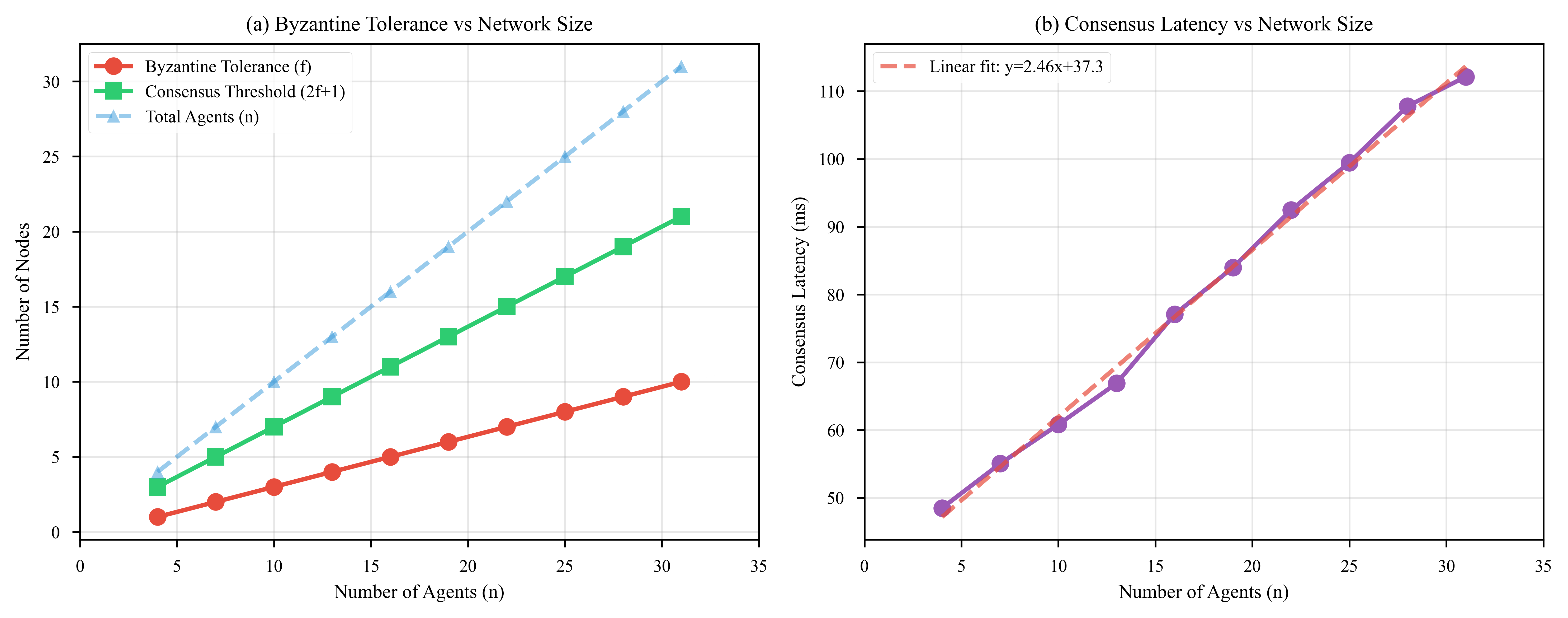}
\caption{Scalability analysis showing (a) Byzantine tolerance and consensus threshold growth with network size, and (b) linear consensus latency scaling for networks up to 31 agents.}
\label{fig:scalability}
\end{figure}

We analyzed theoretical and empirical scalability as shown in Figure \ref{fig:scalability}:

\textbf{Message Complexity:} $O(n \cdot k \cdot h) = O(n)$ for constant $k$ and $h$

\textbf{Vote Collection:} $O(n)$ per consensus round

\textbf{Storage:} $O(m \cdot n)$ for $m$ messages and $n$ agents

The system scales linearly with network size, making it suitable for hospital-scale deployments (10-50 agents). Figure \ref{fig:scalability}(a) demonstrates how Byzantine tolerance $f$ and consensus threshold $2f+1$ grow with network size, while Figure \ref{fig:scalability}(b) shows consensus latency increases linearly at approximately 2.5 ms per additional agent.

\section{Discussion}

\subsection{Security Analysis}

Our system provides several security guarantees:

\textbf{Message Integrity:} SHA-256 signatures ensure tamper detection with probability $1 - 2^{-256}$.

\textbf{Replay Attack Prevention:} Timestamp validation with $\Delta_{max} = 300s$ prevents replay attacks beyond the time window.

\textbf{Byzantine Resilience:} The $2f + 1$ threshold ensures consensus validity with up to $f$ Byzantine nodes, proven by Lamport et al. \cite{lamport1982byzantine}.

\textbf{Sybil Attack Resistance:} Agent identifiers are pre-registered, preventing Sybil attacks in permissioned deployments.

\subsection{Limitations}

Several limitations warrant discussion:

\textbf{Scalability:} While theoretically linear, practical deployments beyond large number of agents may require hierarchical consensus structures.

\textbf{Network Partitions:} The current implementation assumes eventual connectivity. Network partitions could prevent consensus, requiring partition-tolerant extensions.

\textbf{Synchrony Assumptions:} We assume partial synchrony with bounded message delays. Fully asynchronous environments would require different consensus mechanisms \cite{fischer1985impossibility}.

\textbf{Byzantine Threshold:} The $f < n/3$ bound is fundamental to Byzantine consensus. Higher Byzantine ratios require alternative approaches like proof-of-work.

\subsection{Clinical Implications}

The system enables several clinical applications:

\textbf{Collaborative Diagnosis:} Multiple AI agents can validate diagnoses, reducing errors through consensus.

\textbf{Treatment Verification:} Treatment plans undergo Byzantine validation, ensuring safety even if individual agents are compromised.

\textbf{Emergency Coordination:} Fault-tolerant emergency alerts ensure critical messages reach all agents despite failures.

\textbf{Audit Trails:} Cryptographic signatures and consensus records provide immutable audit trails for regulatory compliance.

\subsection{Future Directions}

Several extensions could enhance the system:

\textbf{Adaptive Byzantine Tolerance:} Dynamic adjustment of $f$ based on observed network behavior.

\textbf{Privacy-Preserving Consensus:} Integration of secure multi-party computation for privacy-sensitive medical data.

\textbf{Hierarchical Consensus:} Multi-level consensus for large-scale hospital networks.

\textbf{Machine Learning Integration:} Byzantine-robust federated learning for collaborative model training.

\textbf{Formal Verification:} Formal proofs of consensus properties using theorem provers.

\section{Conclusion}

This paper presented a Byzantine fault-tolerant multi-agent system for healthcare, integrating gossip-based message propagation with cryptographic validation. Our system achieves 100\% consensus accuracy while tolerating up to 33\% Byzantine nodes, demonstrating practical viability for clinical applications.

The key contributions include: (1) a novel integration of Byzantine consensus with healthcare-specific message validation, (2) specialized AI agents with domain-specific tools coordinated through fault-tolerant protocols, (3) gossip-based decentralized message propagation achieving full network coverage, and (4) comprehensive real-time monitoring of consensus operations.

Experimental results validate the system's effectiveness, with mean consensus latency of 45.7 ms and 100\% Byzantine fault detection. The linear scalability makes the system suitable for hospital-scale deployments.

Future work will explore adaptive Byzantine tolerance, privacy-preserving consensus mechanisms, and formal verification of consensus properties. As healthcare Generative AI agentic systems become increasingly distributed and autonomous, Byzantine fault tolerance will be essential for ensuring safety, reliability, and trustworthiness in clinical decision-making.


\begin{thebibliography}{00}

\bibitem{wooldridge2009introduction} M. Wooldridge, \textit{An Introduction to MultiAgent Systems}, 2nd ed. Wiley Publishing, 2009.

\bibitem{shortliffe2012computer} E. H. Shortliffe and J. J. Cimino, \textit{Biomedical Informatics: Computer Applications in Health Care and Biomedicine}, 4th ed. Springer, 2012.

\bibitem{lamport1982byzantine} L. Lamport, R. Shostak, and M. Pease, ``The Byzantine Generals Problem,'' \textit{ACM Transactions on Programming Languages and Systems}, vol. 4, no. 3, pp. 382-401, July 1982.

\bibitem{castro1999practical} M. Castro and B. Liskov, ``Practical Byzantine Fault Tolerance,'' in \textit{Proceedings of the Third Symposium on Operating Systems Design and Implementation (OSDI)}, 1999, pp. 173-186.

\bibitem{zhang2018blockchain} P. Zhang, J. White, D. C. Schmidt, G. Lenz, and S. T. Rosenbloom, ``FHIRChain: Applying Blockchain to Securely and Scalably Share Clinical Data,'' \textit{Computational and Structural Biotechnology Journal}, vol. 16, pp. 267-278, 2018.

\bibitem{ongaro2014search} D. Ongaro and J. Ousterhout, ``In Search of an Understandable Consensus Algorithm,'' in \textit{Proceedings of the 2014 USENIX Annual Technical Conference}, 2014, pp. 305-319.

\bibitem{nakamoto2008bitcoin} S. Nakamoto, ``Bitcoin: A Peer-to-Peer Electronic Cash System,'' 2008. [Online]. Available: https://bitcoin.org/bitcoin.pdf

\bibitem{demers1987epidemic} A. Demers, D. Greene, C. Hauser, W. Irish, J. Larson, S. Shenker, H. Sturgis, D. Swinehart, and D. Terry, ``Epidemic Algorithms for Replicated Database Maintenance,'' in \textit{Proceedings of the Sixth Annual ACM Symposium on Principles of Distributed Computing}, 1987, pp. 1-12.

\bibitem{jelasity2007gossip} M. Jelasity, S. Voulgaris, R. Guerraoui, A.-M. Kermarrec, and M. van Steen, ``Gossip-based Peer Sampling,'' \textit{ACM Transactions on Computer Systems}, vol. 25, no. 3, Aug. 2007.

\bibitem{malkhi2005concise} D. Malkhi, Y. Mansour, and M. K. Reiter, ``On Diffusing Updates in a Byzantine Environment,'' in \textit{Proceedings of the 18th IEEE Symposium on Reliable Distributed Systems}, 1999, pp. 134-143.

\bibitem{isern2010agent} D. Isern, D. Sánchez, and A. Moreno, ``Agents Applied in Health Care: A Review,'' \textit{International Journal of Medical Informatics}, vol. 79, no. 3, pp. 145-166, 2010.

\bibitem{corchado2008healthcare} J. M. Corchado, J. Bajo, and A. Abraham, ``GerAmi: Improving Healthcare Delivery in Geriatric Residences,'' \textit{IEEE Intelligent Systems}, vol. 23, no. 2, pp. 19-25, 2008.

\bibitem{decker1996designing} K. Decker and J. Li, ``Coordinated Hospital Patient Scheduling,'' in \textit{Proceedings of the Third International Conference on Multi-Agent Systems}, 1998, pp. 104-111.

\bibitem{nguyen2019federated} D. C. Nguyen, P. N. Pathirana, M. Ding, and A. Seneviratne, ``Blockchain for Secure EHRs Sharing of Mobile Cloud Based E-Health Systems,'' \textit{IEEE Access}, vol. 7, pp. 66792-66806, 2019.

\bibitem{fischer1985impossibility} M. J. Fischer, N. A. Lynch, and M. S. Paterson, ``Impossibility of Distributed Consensus with One Faulty Process,'' \textit{Journal of the ACM}, vol. 32, no. 2, pp. 374-382, Apr. 1985.

\end{thebibliography}
\end{document}